\documentclass[sigconf]{acmart}
\pdfoutput=1 % Required for ArXiv submission
\renewcommand\footnotetextcopyrightpermission[1]{} % removes footnote with conference information in first column
\usepackage{booktabs} % Importing table commands, such as toprule and midrule

\begin{document}

\title{Real-Time Detection of Dictionary DGA Network Traffic using Deep Learning}

\author{Kate Highnam}
\affiliation{
  \institution{Imperial College London}
}
\email{k.highnam19@imperial.ac.uk}
%%%%%%%%%%%%%%%%%%%%%%%%%%%%%%%%%%%%%%%%%%%%%%%%%%%%%%%%%%%%%%
\author{Domenic Puzio}
\affiliation{
  \institution{Kensho Technologies}
}
\email{domenicvpuzio@gmail.com}
%%%%%%%%%%%%%%%%%%%%%%%%%%%%%%%%%%%%%%%%%%%%%%%%%%%%%%%%%%%%%%
\author{Song Luo}
\affiliation{%
  \institution{Tencent}
}
\email{luosong@gmail.com}
%%%%%%%%%%%%%%%%%%%%%%%%%%%%%%%%%%%%%%%%%%%%%%%%%%%%%%%%%%%%%%
\author{Nicholas R. Jennings}
\affiliation{%
  \institution{Imperial College London}
}
\email{n.jennings@imperial.ac.uk}

\renewcommand{\shortauthors}{Highnam, et al.}

\begin{abstract}
  Botnets and malware continue to avoid detection by static rules engines when using domain generation algorithms (DGAs) % plohmann 2016
  for callouts to unique, dynamically generated web addresses. Common DGA detection techniques fail to reliably detect DGA variants that combine random dictionary words to create domain names that closely mirror legitimate domains. To combat this, we created a novel hybrid neural network, Bilbo the “bagging” model, that analyses domains and scores the likelihood they are generated by such algorithms and therefore are potentially malicious. Bilbo is the first parallel usage of a convolutional neural network (CNN) and a long short-term memory (LSTM) network for DGA detection. Our unique architecture is found to be the most consistent in performance in terms of AUC, $F_1$ score, and accuracy when generalising across different dictionary DGA classification tasks compared to current state-of-the-art deep learning architectures. We validate using reverse-engineered dictionary DGA domains and detail our real-time implementation strategy for scoring real-world network logs within a large enterprise
  \footnote{
  This work was done while working at 
%   a large financial institution 
%   FIXME ANONYMOUS SUBMISSION, suppressing this information...
    Capital One in 2017-18 and is based on public talks we gave in 2018. 
%   Further experiments and comparisons have been added since to cover the vast improvements within the field in the last year.
  }. 
  In four hours of actual network traffic, the model discovered at least five potential command-and-control networks that commercial vendor tools did not flag. 
\end{abstract}



\keywords{domain generation algorithm, deep learning, malware, botnets, network security, neural networks}

\maketitle

\section{Introduction}
Malware continues to pose a serious threat to individuals and corporations alike \cite{plohmann2016comprehensive}. Typical attack methods such as viruses, phishing emails, and worms attempt to retrieve private user data, destroy systems, or start unwanted programs. The majority of these attacks may be launched through the network \cite{oprea2015enterpriselogdata}, posing a major threat to any Internet-facing device. Some malware reaches out to a command and control (C\&C) centre hosted behind domains generated by an algorithm (DGA domains) after it infiltrates the target system to receive further instructions. Identification of such domains in network traffic allows for the detection of malware-infected machines. 

A single active DGA has been seen generating up to a few hundred domains per day \cite{plohmann2016comprehensive}. At scale within a company, this is infeasible for a human analyst to triage amidst the thousands of benign domains occurring simultaneously. Automated detection systems are developing but the sightings of DGAs in worms, botnets, and other malicious settings is growing \cite{unit422019dga}.

In addition, malware that employs DGAs intentionally obfuscates its network communication by utilising random seeds when generating their domains  \cite{domainz, yadav2010detecting, oprea2015enterpriselogdata, plohmann2016comprehensive, unit422019dga}. Most known DGAs combine randomly-selected characters like ``myypqmvzkgnrf[.]com'', ``otopshphtnhml[.]net'', and ``uqhucsontf[.]com'' \footnote{For the rest of this paper, all domain URLs will be referred to with [.] to prevent automatically assigning these domains as real URLs one might click.}. 

However, DGAs that combine random words from a dictionary like ``milkdustbadliterally[.]com'', ``couragenearest[.]net'', and ``boredlaptopattorney[.]ru'' \cite{dgarchive} are meaningfully harder for humans to detect (see Table \ref{fig:dga-examples} for comparison). In this paper, we will refer to this type of DGA as a \textit{dictionary DGA} and focus on those using dictionaries composed of English words. 

Common defences against malicious DGA domains include blacklists \cite{leverlustrum,kuhrer2014paint}, random forest classifiers \cite{ahluwalia2017detecting,yu2017inline,woodbridge2016lstm}, and clustering techniques \cite{dnsclusteringdga, pereira2018wordgraph}.
When the lists are well maintained and the features are chosen carefully, these methods have acceptable efficacy. However, both blacklists and these models possess serious limitations: relying on hand-picked features which are time-consuming to develop, lacking the ability to generalise with the few manual features implemented, and requiring continuous expert maintenance. More comprehensive tactics are necessary to detect incessant new DGAs stemming from network-based malware.

Recent innovations using deep learning have state-of-the-art accuracy on DGA detection. Such models are highly flexible with the proven success in  complex language problems. They do not require hand-crafted features that are time-intensive to make and easy to evade. Woodbridge et al. \cite{woodbridge2016lstm} were the first to present a long short-term memory (LSTM) network for DGA classification. Other architectures were later applied, such as further variations on an LSTM \cite{tran2018lstm, akarsh2019deep, vinayakumar2018scale, yu2017inline, lison2017automatic},
a convolutional neural network (CNN) \cite{saxe2017expose, zhou2019cnn},
and a hybrid CNN-LSTM model \cite{yu2018character}. Although successful for random-character DGA domains, these classifiers have largely been ineffective in identifying dictionary DGA domains.
% since their algorithms generate domains that mirror the character distribution of actual domains \cite{raghuram2014unsupervised}.
These models also perform well on their various testing sets but their performance can suffer when attempting to generalise to new DGA families or new versions of previously seen families.

Against this background, we present a novel deep learning model for dictionary DGA detection. This advances the state of the art in the following ways. First, we present the first usage of parallel CNN and LSTM hybrid for DGA detection, specifically applied to dictionary DGA detection. The model is trained on standard large-scale datasets of reverse-engineered dictionary DGA domains. It achieves the most consistent success at dictionary DGA classification amongst state-of-the-art deep learning architectures for classification, generalisability, and time-based resiliency. Second, we detail our insights into dictionary DGA domains' inter-relationships and their effect on generalisability of models as an outcome. Third, we validate our model on live network traffic in a large financial institution. In four hours of logs, it discovered five potential C\&C networks that commercial vendor tools did not flag. Finally, we detail our scalable implementation strategy within the security context of a corporation for real-time analysis.

\section{Background}

An ever-growing number of malware rely on communication with C\&C channels to receive instructions and system-specific code \cite{plohmann2016comprehensive}.
The destination (domain or IP address) of this channel can be hard-coded in the malware itself, making its location discoverable via reverse engineering or straightforward log aggregation techniques. Once known, this domain or IP address can be blacklisted, rendering the malware inert. 
To avoid this single point of failure, malware authors employ domain fluxing, in which the destination of the C\&C communication changes systematically as the attacker registers new domains to the C\&C hub.

The key to malware domain fluxing is the use of unique and likely unregistered domains that are known to the attacker but can blend in to regular traffic. To accomplish this, malware families employ domain generation algorithms (DGAs) to create pseudo-random domains for use in communication. These domains are used for short periods of time and then phased out for newly-generated domains; this quick turnover means that manual techniques are not effective. Additionally, reverse engineering these algorithms may be slow or impossible if the malware is encrypted. For the vast majority of malware samples, traffic related to malicious activity is present in networks weeks or months before the malware is analysed and blacklisted \cite{leverlustrum}.

% FIXME(Should I include this?) 
To prevent DGA-based malware from exfiltrating, disabling, or tampering with assets, institutions must detect malicious traffic as soon as possible. Throughout this paper we will discuss our solution while keeping in mind that it must be practical, operating in real-time, enriching contextual data within in true threat environments.

\begin{table}[t]
 \caption{  
Examples of domains from our training data, comprised of domains from the Alexa Top 1 Million list and domains generated by dictionary-based DGA families (discovered through reverse engineering) from DGArchive.
\label{fig:dga-examples}
}
\begin{center}
\begin{tabular} [width=0.9\columnwidth] { c  c}
  \toprule
	\textbf{Legitimate} & \textbf{Malicious} \\ 
  \midrule
    \texttt{microsoft} & \texttt{lookhurt} \\
    \texttt{linkedin} & \texttt{threetold} \\
    \texttt{paypal} & \texttt{threewear} \\
    \texttt{steamcommunity} & \texttt{pielivingbytes} \\
    \texttt{dailymotion} & \texttt{awardsbookcasio} \\
    \texttt{stackoverflow} & \texttt{blanketcontent} \\
    % \texttt{nytimes} & \texttt{bottomconcerned} \\
    \texttt{facebook} & \texttt{degreeblindagent} \\
    % \texttt{whatsapp} & \texttt{lineknitcalculators} \\
    \texttt{soundcloud} & \texttt{mistakelivegarage} \\
  \bottomrule
\end{tabular}
    % \vspace{-1em}
\end{center}
\end{table}

\subsection{Domain Generation Algorithms (DGAs) }

DGA usage spans a variety of cases, from 
%advertisement networks seeking to avoid filtering 
benign resource generation to phishing campaigns and the management of botnets, groups of machines that have been infected by malware, such as Kraken \cite{royal2008analysis}, Conficker \cite{porras2009analysis,porras2009inside}, Murofet \cite{shevchenko2010domain}, and others \cite{yadav2012detecting}. 
 The goal of all DGAs is to generate domains that do not already exist and, for malicious cases, will not be flagged by vendor security tools or analysts. To accomplish this, DGA authors typically use either character-based or dictionary-based pseudo-random assembly process to form domains.
 
% In the case of character-based DGAs from malware families such as \texttt{tinba} and \texttt{virut}, domains like ``ubzeaq[.]com'' and ``otopshphtnhml[.]net'' are generated. 

Each method has benefits and downfalls. Character-based DGA domains are more likely to not be registered. But to a human security analyst, gibberish domains made from character-based DGAs stand out from human-crafted domains due to their phonetic implausibility and lack of known words within them. There is a visible unique pattern underlying character DGA domains, such as ``lrluiologistbikerepil'', that dictionary DGA domains, like ``recordkidneyestablishmen'', do not follow. Dictionary DGA domains are more challenging to detect when scanning logs because they are pronounceable, contain known words, and mirror the character distribution of legitimate English domains \cite{anderson2016deepdga}. See similarities between known dictionary DGA domains and benign domains in Table \ref{fig:dga-examples}.
%Such domains also differ from legitimate domains in their character distributions, since all letters are equally common when chosen at random. 

DGA detection systems have been implemented to assist in highlighting DGA domains for further investigation. These have largely been tailored towards character-based DGAs. Character-based DGAs are more common: of 43 known reverse-engineered DGAs available in DGArchive \cite{dgarchive}, 40 of them use a seed to pseudo-randomly assemble characters or a word surrounded by random characters to form a domain name. Most methods for generic DGA analysis still struggle to identify dictionary-based DGA malware families because they classify all DGAs rather than focusing on specific algorithms. 

% A variety of tactics have attempted to address these kinds of domains, such as trying to learn the words used by the dictionary DGA to see how the algorithm is generating these domains, but it has been shown that DGAs evolve over time \cite{pereira2018wordgraph}. Thus, proposed solutions must be able to generalise over changing of words for multiple dictionaries and alterations in algorithms that could generate them.
This paper will focus on classifying the largest available sets of known dictionary DGA domains: \texttt{gozi} \cite{rovnixbot2014gozi}, \texttt{matsnu} \cite{skuratovich2015matsnu}, and \texttt{suppobox} \cite{geffner2013end}. Each varies in the dictionary-based domain generation tactic, the length of the domain, and the dictionary corpus. 
% Unlike other DGAs, all three consistently uses word lists to generate domains.
These dictionary DGA families are often undetected by methods proposed in prior research aimed at general DGA detection because of the large number of families available for other types of DGAs. By targeting where others are weaker, our model can provide greater coverage when used in conjunction with generic DGA models and other contextual information for increased confidence in identification.

Much of prior DGA research has involved making lookups into historical or related domain name server (DNS) records. Such methods often rely on signals attained from Non-Existent Domain (NXDomain) responses when unregistered domains are queried. Since DGAs often generate hundreds of domains per day and at most only a few of those domains are actually registered by the attacker, large numbers of these requests result in NXDomains. Many NXDomain responses from the same computer are unlikely to result from expected user behaviour, and thus this pattern of DNS traffic can be associated with DGA activity \cite{dnsclusteringdga, yadav2012detecting, krishnan2013crossing}. 

However, such queries within high-volume DNS log data can be prohibitively slow and unsuitable for real-time decision-making needed to reduce the risk of compromise. It is for this reason that our model considers limited data, only the domain name, rather than all of the potential fields given through standard network logging. We also only use open source datasets rather than restricted NXDomain lists for reproducibility and to provide an accessible starting point for others looking to tailor this system to their own environment.

\subsection{Related Work} \label{relatedwork}

% Malicious URL analysis has been a growing issue in recent years and addressed in a variety of ways. Initial motivation included identifying phishing emails \cite{verma2015character, bahnsen2017classifying} and flagging DGA domains of various malware or botnets \cite{plohmann2016comprehensive, porras2009inside}. As threats using DGAs grew and the limitations of blacklisting were shown, more focus was drawn to DGA analysis \cite{kuhrer2014paint, geffner2013end, leverlustrum}.

Defensive tactics began analysing network logs with statistical or manually selected features instead of static blacklisting or rules when it became overwhelming to maintain them. Unsupervised probabilistic filtering \cite{raghuram2014unsupervised} and random forest models \cite{schiavoni2014phoenix, ahluwalia2017detecting} were some of the leading systems for detecting DGAs. 

Future techniques included more contextual information which improved the longevity of detection systems. Clustering \cite{yadav2010detecting, yadav2012detecting, zhou2013dga}, Hidden Markov Models (HMMs) \cite{dnsclusteringdga}, random forests models \cite{schuppen2018fanci, verma2015character, yang2018feature}, and sequential hypothesis testing \cite{krishnan2013crossing} used data such as WHOIS or NXDomain responses with the domain to identify DGAs. However, a number of these techniques require batches of live data to maintain relevancy or high volumes of data which are not typically feasible in real-time environments.

Deep learning first addressed DGA detection with work by Woodbridge et al. \cite{woodbridge2016lstm}, an implementation of an LSTM used for nonspecific DGA analysis. Their experiments show that their deep learning approach, an LSTM network, outperforms a character-level HMM and a random forest model that utilise features such as the entropy of character distribution. Their analysis and implementation led to a large success for identifying most DGA families; however, their LSTM did not score highly on \texttt{suppobox} or \texttt{matsnu}, dictionary DGA families. 
% Woodbridge et al. leave in-depth analysis of these families to future work. 

Since then others have joined the field, implementing a variety of deep learning models. Several took the LSTM model from Woodbridge et al. %\cite{woodbridge2016lstm} 
and provided improvements. Tran et al. \cite{tran2018lstm} took the native class imbalance of DGA data into account. Others updated the training data with other known DGA datasets \cite{akarsh2019deep, yu2017inline} or added more contextual information to the score \cite{curtin2019smashword}. Another altered the original architecture of their LSTM to a bi-directional LSTM layer \cite{lison2017automatic}, demonstrating the potential enhancements of changing the model's architecture.

When a CNN was applied to text classification \cite{johnson2014effective, kim2014convolutional, zhang2015character} and showed success over an LSTM on some tasks \cite{yin2017comparative, zhou2019cnn}, it was eventually applied to malicious URL analysis \cite{saxe2017expose}. Other approaches to this problem include a Generative Adversarial Network (GAN), showing that the arms race for DGA detection could advance on its own \cite{anderson2016deepdga}. Recent work combining CNN convolutions and LSTM temporal processing into new sequential hybrid models have also been brought to this problem \cite{chen2017ensemble, kim2016character, mohan2018spoof, yu2018character}.
Other comparative works have been published attempting to finalise which model is the best for DGA detection \cite{mac2017dga, berman2019survey, sivaguru2018evaluation, yu2018character, feng2017classification, yu2017inline}. Their evaluations state deep learning maintains greater success over random forest models trained using manually-selected features, but do not consider the greater context of the model's deployment or implementation environment. Our research picks up this work, systematically evaluating deep learning architectures to specifically target where most DGA detection systems consistently underperform: dictionary DGAs. 

Koh et al. were one of the first to train deep learning to specifically target dictionary DGA domains \cite{koh2018lstmembed}. Utilising a pre-trained embedding for the words within the domain, they trained an LSTM both on single-DGA and multiple-DGA data sets. While their results set the bar for dictionary DGA detection, their model had severe limitations from its context-sensitive word embedding on what it could learn and they did not use all available data during training and testing. Another related work on dictionary DGA detection is WordGraph  from Pereira et al. \cite{pereira2018wordgraph}. They take large batches of NXDomains and the longest common substring (LCS) of every pair within the set, connecting any co-occurring LCS within a single domain name to construct their WordGraph. The dictionary DGA domains are shown to cluster whereas benign domains have no discernible pattern and is shown to generalise over changes to the DGA's dictionary. A random forest classifier is trained on the patterns between domains to identify dictionary DGA patterns. This method shows promise at adapting to different DGAs. However, it is too computationally intensive for many systems to support for only domain name analysis.

\subsection{Real-Time Deployment Environment}

Within a large corporation with thousands of employees, security tools struggle to assist analysts attempting to monitor corporate assets. Analysts investigating anomalous activity use a variety of filters to limit the data they need to consider before finalising a verdict on any given activity. We assume other filters for response type, network protocol, NXDomain results, proxy labels, etc. are also included. Scores from a model for dictionary DGA detection would be added into the system for analysts to include whichever additional information they deem necessary.

Much like the work by Kumar et al. \cite{kumar2019deploymentsystem} and Vinayakumar et al. \cite{vinayakumar2018scale}, we aim to not only address this cyber security issue with text classification techniques, but also the greater system in which the model would be deployed. Prior systems consider the various model performance metrics on common data sets as well as the real-world generalisability, response time, and scalability of their chosen model when scoring domains in real time. We extend their work to new controlled tests and describe deploying detection systems within a corporate environment.

% The authors suggest that the failings of detecting dictionary DGAs could be due to sample bias that favors character-based DGA families over dictionary-based DGA families. 

%%%%%%%%%%%%%%%%%%%%%%%%%%%%%%%%%%%%%
\begin{figure}[th]
\begin{center} % FIXME Fix diagram
    \includegraphics[width=0.9\columnwidth]{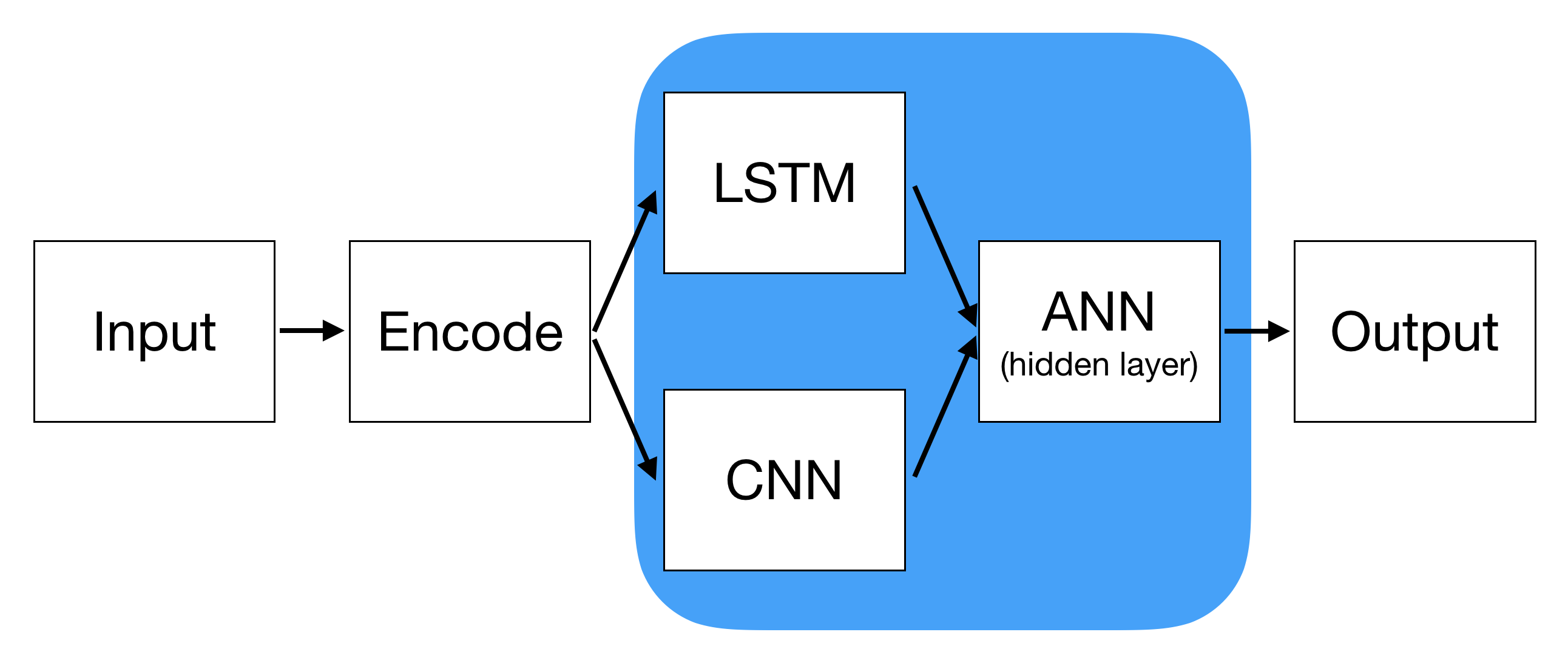}
	\caption{
    High-level architecture of Bilbo; the component models are highlighted in blue. Raw domains are input and encoded into sequences before being passed to the separate LSTM and CNN architectures. The features extracted by each of these component architectures are sent to a single layer ANN or a hidden layer, which is then flattened to produce the output, a single score. 
    \vspace{-1em}
    \label{fig:hybrid-arch}
  	}
\end{center}
\end{figure}
%%%%%%%%%%%%%%%%%%%%%%%%%%%%%%%%%%%%%

\section{Bilbo the "Bagging" Hybrid Model}

We present a new deep learning model to deploy for real-time dictionary DGA classification.
As mentioned before, deep learning architectures are capable of learning variations to dictionaries and DGAs, with the added benefit of training quickly. There have been many deep learning architectures published for this task for state-of-the-art comparison.

% We present the first parallel usage of a CNN and LSTM within a hybrid model for DGA detection, targeting dictionary DGA detection. This novel bagging architecture analyzes domains from two perspectives, resulting in state-of-the-art generalization for this task.
%Its novel architecture allows it to specialize in two different ways of learning trends within the samples and can generalize for this task better than any model published. 
% In this section, we describe the motivations behind each component within our model before showing evidence of its success in Section 6.
% \textit{Briefly explain the benefits and reasoning behind a hybrid model. Why is it helpful? What are the alternatives? What is the norm when using them?}

% \subsection{LSTM}
% A long short-term memory (LSTM) network \cite{og-lstm} is a generalization of a recurrent neural network (RNN), which was built for processing sequential data. 
Since we can treat domains as sequences of characters, LSTM  models are a natural fit for classifying DGA domains. LSTM nodes make decisions about one element in the sequence based on what it has seen earlier in the sequence. Thus, LSTM nodes learn parameters that are shared across the elements of sequence. This parameter sharing allows LSTMs to scale to handle much longer sequences than would be practical for traditional feedforward neural networks \cite{goodfellow2016deep}.
% However, RNNs struggle to deal with long-term sequential dependencies, features from the beginning of the sequence that might be relevant for prediction on the later elements of the sequence. Due to vanishing gradients, the weights given to these long-term dependencies become exponentially smaller than those for short-term dependencies. To solve this problem, LSTMs were introduced \cite{goodfellow2016deep}. 
% LSTM nodes maintain a state that can be forgotten, remembered, or changed with each new input. 
For example, an LSTM neuron might recall that it has seen seven vowels in a nine-character domain, making it unlikely that the domain is made up of natural English text. 
% LSTMs have proven successful in a wide variety of text and language problems, including domain classification \cite{woodbridge2016lstm}. 
This sequential specialisation of LSTMs attracted us initially, but we found it alone could not generalise to new dictionary DGAs as well as other architectures.

% \subsection{CNN}
Others have applied CNNs in various forms since used for URL analysis by Saxe et al. \cite{saxe2017expose}.
Convolutional neural networks (CNNs) were designed to handle information that is in a grid format, such as pixels in an image matrix. 
% CNNs have demonstrated state-of-the-art performance on image processing tasks, since, unlike feedforward networks, they can compute over large grids by applying a filter to an image and intelligently extracting certain attributes. 
By treating text as a one-dimensional grid of letters, CNNs were shown to have excellent results for natural language tasks \cite{kim2014convolutional, zhang2015character}. We translate domain names to arrays of characters, allowing the CNN to examine local relationships between characters via a sliding window, thus grouping characters together into words. For example, the domain ``facebook'' can be broken down into four-character windows: ``face'', ``aceb'', ``cebo'', ``eboo'', and ``book''. By dividing character arrays into smaller, related parts in this manner, CNNs demonstrated success on URL classification tasks \cite{saxe2017expose}.

% \subsection{Hybrid Model}

% With two models proven to perform well on text classifications, 
When multiple models perform well on the same task,
many practitioners have combined models or model architectures to enhance the various benefits they individually provide. The most common technique is to combine pre-trained models to form an ensemble model, where each individual model produces a score and these scores are combined in some way to produce a new score. In this context we could train a general DGA classifier that combines one model trained to classify character DGA domains and another trained to classify dictionary DGA domains. The benefit of combining both models is dependent on how they are combined and how it decides which model to ``trust'' for its final decision without the context of how they were developed.

Hybrid models are similar to ensemble models, but rather than taking the individual score from each component, a hybrid model combines the architectures before the extracted features are reduced to a single score. These models are trained as a single end-to-end model. A hybrid architecture allows the model to learn which combinations of features of the input are significant indicators for accurate classification. Most common hybrid models combine architectures by stacking them in different ways. For instance, using a CNN's convolutional layer to extract features and then feed them into an LSTM layer
\cite{vosoughi2016tweet2vec, chen2017ensemble, kim2016character, mohan2018spoof, yu2018character}.

Our novel hybrid model, as seen in Figure \ref{fig:hybrid-arch}, processes domain names via an LSTM layer and a CNN layer in parallel. The outputs of these two architectures are then aggregated or ``bagged'' by a single-layer ANN. This ``bagging'' is a vital opportunity for this model to discern which parts of the captured information from the LSTM and CNN assists the best when labelling dictionary DGA and benign domains. Inserting an ANN instead of a single function increases the potential optimisation of the ``bagging''.
Because of the importance of this piece in the architecture, we named our model Bilbo the "bagging" model. 
Unlike ensembles which optimise its components prior to conjoining, hybrids optimise over all the components.
As demonstrated in our results (Section \ref{Results}), Bilbo successfully combines LSTM, CNN, and ANN layers for dictionary DGA detection and is the best at consistently classifying dictionary DGAs amongst state-of-the-art deep learning models.

% Another way to form a hybrid model is by feeding inputs into two or more model architectures in parallel. An LSTM and a CNN could both receive the domain name and the features they extract in their hidden layers that would normally be flattened to a single score could instead be fed into a single-layer ANN. 
% The entire model would be trained as one. This architecture can be viewed as, in a way, ``bagging'' the features from the sub-architectures. Thus creating Bilbo the "bagging" model. 

%%%%%%%%%%%%%%%%%%%%%%%%%%%%%%%%%%%%%
\begin{figure}[th]
\begin{center}
    \includegraphics[width=0.9\columnwidth]{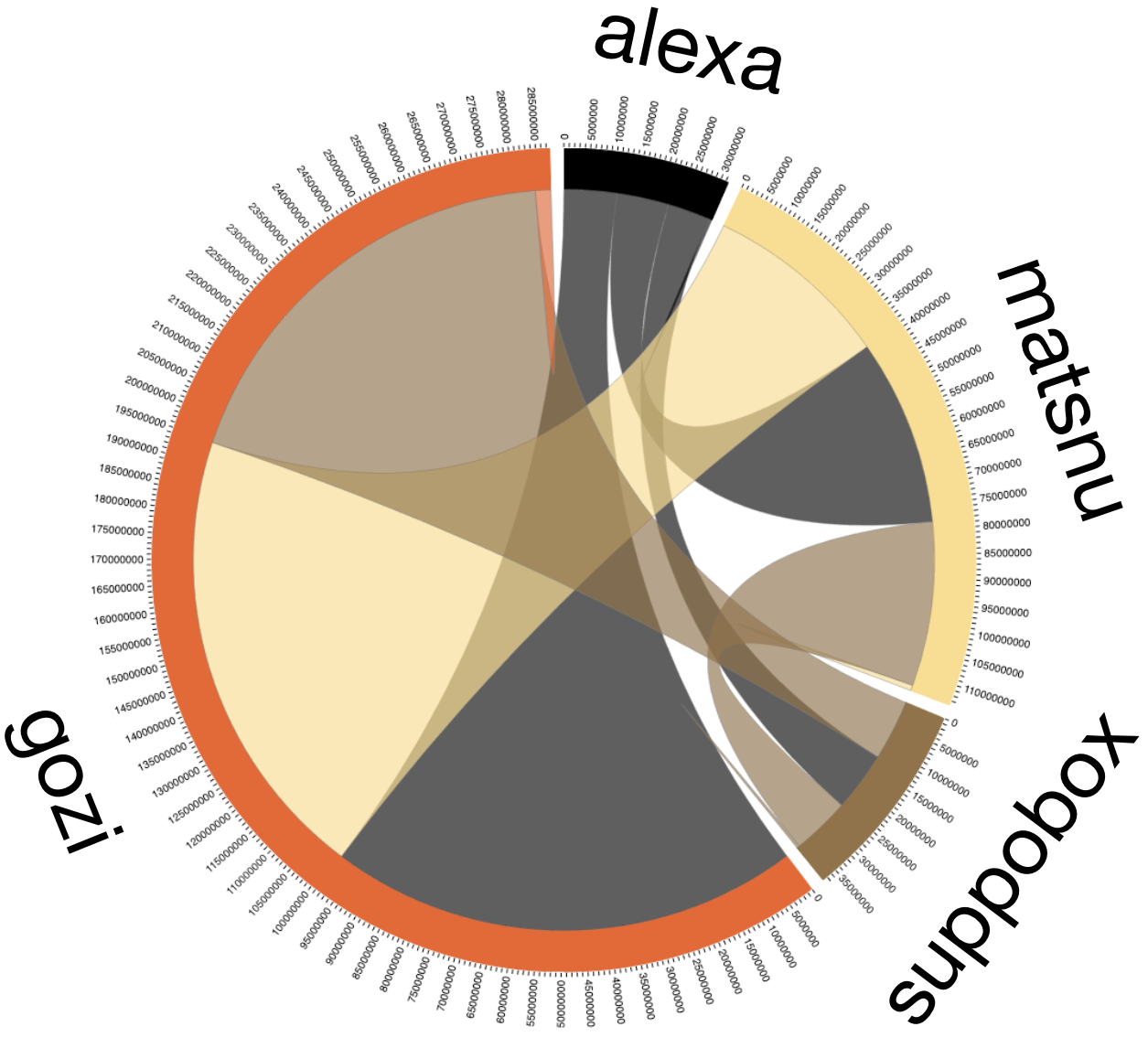}
	\caption{
    Comparing the shared largest common substrings from within each domain family considered during our classification  (\texttt{alexa}, \texttt{suppobox}, \texttt{gozi}, and \texttt{matsnu}). The circumference is grouped by colour for each family. The counts are for the number of times the overlapping LCS was an LCS for a domain pair within a given family. Note that any overlap in the centre has no meaning and the counts contain overlap between LCS shared between one pair of families and any other.
    \vspace{-1em}
    \label{fig:chord-counts}
  	}
\end{center}
\end{figure} 
%%%%%%%%%%%%%%%%%%%%%%%%%%%%%%%%%%%%%

\section{Data Analysis} \label{briefdataanalysis}

% We only consider the domain names from DNS logs to limit the amount of data necessary for the model to process. Also, we understand the model is unable to catch compromised websites which might be evident from URL extensions. This is a known assumption we inform analysts of when they use dictionary DGA detection scores. 
% To allow for reproducibility and provide an accessible starting point for others looking to tailor this detection system to their own environment, we used two large labeled open-source datasets.
% to train, test, and validate our model. % FIXME phrasing

To better understand the success and failures of the models used in our tests, we conducted a brief analysis of our data set of known dictionary DGAs.
The dictionary DGA domains were selected from collections of related DGAs, called DGA families, published on DGArchive \cite{dgarchive}, a trusted database of domains extracted from reverse-engineered DGA malware. 
From this source, several families of DGAs were empirically identified as solely dictionary DGAs based on the structure of the domain names generated by malware samples. 
The families selected were \texttt{suppobox}, \texttt{gozi}, and \texttt{matsnu} with domains collected over two years (2016-17) by DGArchive. 
After removing duplicate domain names, the resulting selection contained $137,745$ samples of \texttt{suppobox}, $18,539$ samples of \texttt{matsnu}, and $20,313$ samples of \texttt{gozi}.

The legitimate domains in the training set originate from the Alexa Top 1 Million domains, measured in 2016 \cite{alextop1mil}. The Alexa list ranks domains by the number of times each has been accessed.
Since DGA-based malware tends to use domains for short periods of time, we assume that top Alexa domain names are human-generated and label them as non-DGA. These popular domains, mostly containing valid English words, encourage the model to learn characteristics of legitimate combinations of English words. 
We randomly sampled an equivalent number of domains from Alexa to match the total number of dictionary DGA samples available. 

To further understand our data, we conducted several comparisons:

\begin{enumerate}
  \item By extracting the longest common substrings within each family, compare the lists between families for dictionary similarity. See Figure \ref{fig:chord-counts} for a summary of those results
  \item Using the widely adapted Jaro-Winkler algorithm for string similarity \cite{gomaa2013jwscore}, we compared every domain in our data set within their own families and with every other family. The histogram in Figure \ref{fig:hist20-testcase3} shows us how similar families are and how this could influence the results for generalisability. 
\end{enumerate}

\subsection{Longest Common Substring (LCS)}

The application of this algorithm was inspired by Pereira et al.'s technique for dictionary extraction \cite{pereira2018wordgraph}. We applied this to each individual group (\texttt{alexa}, \texttt{suppobox}, \texttt{gozi}, and \texttt{matsnu}) to generate a list of every LCS between pairs of domains. These lists contain all possible dictionary words used to generate the domains. By comparing the lists between the families, we can see how learning one family's list could assist in identifying the other. Figure \ref{fig:chord-counts} visualises the overlap between sets with a chord diagram.

The circumference is partitioned into four parts and is labelled with the count for the number of times overlapping substrings were seen as the LCS for a domain pair within its family.
For instance, look at the black vertical chord between \texttt{gozi} and \texttt{alexa}. The colour black means that \texttt{alexa}, the family assigned black, is the smaller portion of this relationship, i.e. fewer of its LCS (approximately 10 million) are within the overlap with \texttt{gozi} (approximately 100 million).

LCS overlapping between \texttt{alexa} and \texttt{gozi} also include LCS from other overlaps. \texttt{gozi}'s large partition of the circumference while also being the smallest family means it overlaps frequently with other groups. Overall \texttt{matsnu} and \texttt{gozi} have the largest overlap, sharing $8.6\%$ of their LCS and $92\%$ of their LCS when including the number of times it was seen as the LCS of a pair. The longest LCS between them was 14 characters; the average length for LCS was $4.238$ characters. Therefore, there must be only a few very common substrings between the families, which deep learning models could learn.
% The longest LCS between them was 14 characters; the average length for LCS was $4.238$ characters. Therefore, there must be only a few very common substrings between the families, which deep learning models could learn.
% For example, the wider coverage of the partition for \texttt{gozi} mapping to the smaller coverage of \texttt{matsnu} means that many substrings seen frequently in \texttt{gozi} were also present, but in fewer cases, as the longest substring within \texttt{matsnu}. The colour of the arches between sides are coloured whichever family has few LCS overlap in the relationship. Arches with a partition with no other family besides the source mean those were LCS not shared by any two families. 

\subsection{Jaro-Winkler (JW) Score}

\begin{figure}[th]
\begin{center}
    \includegraphics[width=0.9\columnwidth]{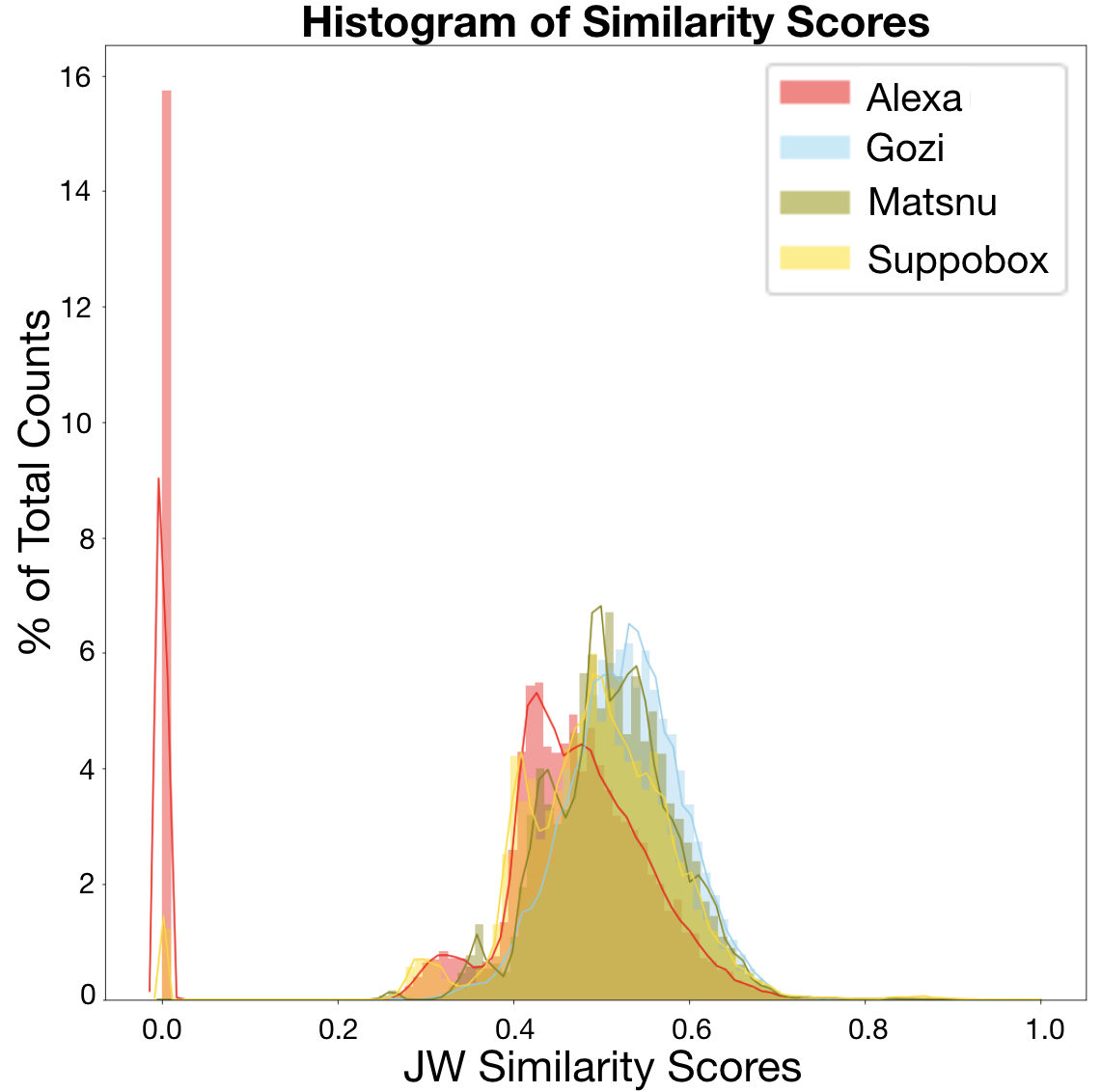}
	\caption{
    Histogram of the Jaro-Winkler scores of each dictionary DGA family and Alexa. A distribution line was drawn over it to assist in tracing the trends of the scores.
    \vspace{-1em}
    \label{fig:hist20-testcase3}
  	}
\end{center}
\end{figure}

To understand the similarity of an entire domain string with any other domain, we used the JW score \cite{gomaa2013jwscore}. This algorithm takes the ordering of the characters and the collection of characters to develop a score between \texttt{[0,1]}. The closer the score is to one, the more similar the domains are to one another. We compared every domain to generate diagrams such as Figure \ref{fig:hist20-testcase3}. 

Most families follow the same distribution with a mean of about $0.5$ for JW score. However, notice the slight skew in \texttt{alexa} and \texttt{suppobox}. Due to a large percentage of their domains having little to no JW similarity, the average score for \texttt{alexa} was $0.4023$ and \texttt{suppobox} was $0.4901$. This slight difference is amplified when considering other aspects of the family. Both \texttt{suppobox} and \texttt{alexa} have the smallest average lengths of domains at $13$ and $9$ characters, respectively. Both groups have a standard deviation of approximately four characters and most frequent length of about eight characters. With this, the low JW scores for \texttt{alexa} and \texttt{suppobox} make sense with shorter domains.

The other sets, \texttt{matsnu} and \texttt{gozi}, are much longer in comparison with most frequent lengths of $14$ and $23$ characters, respectively. The dictionary for their DGAs seems to select from shorter, 3-5-character words. Since there are less possible combinations of valid short words, more overlap between \texttt{gozi} and \texttt{matsnu}, which is also apparent in Figure \ref{fig:chord-counts}.

This exploratory data analysis helped us develop an intuition around how different dictionary DGAs relate to each other and gave us hope that models would pick up on these relationships even though most of these families use different dictionaries and generation algorithms. Also, this same analysis should prove useful when comparing and expanding the model with other dictionary DGA families as they emerge.

%%%%%%%%%%%%%%%%%%%%%%%%%%%%%%%%%%%%%
%%%%%%%%%%%%%%%%%%%%%%%%%%%%%%%%%%%%%

\section{Experimental Design} \label{experimentaldesign}

We frame the DGA detection problem as a binary text classification task on only the domain string. The score provided by our model can then be used independently or be enriched with additional security context.  
In this section, tests are designed on known labelled data to demonstrate the baseline performance of each model. These experiments reflect practitioner concerns on model deployments within real-world context:

\begin{enumerate}
  \item Testing the model's ability to do binary classification with of benign and dictionary DGA domains
  \item Evaluating the model's generalisability for identifying unseen dictionary DGAs 
%   from training on some of the dictionary DGA families to testing on a new one
  \item Examining the model's scores as the dictionaries and DGAs evolve over time, how well can the model classify new dictionary DGA domains from known families
\end{enumerate}

We compare Bilbo to four deep learning models: a single-layer ANN, CNN, LSTM, and MIT's CNN-LSTM Hybrid \cite{vosoughi2016tweet2vec, yu2018character}. 
Each is based off of state-of-the-art models for DGA classification; the implementation for each is described below. 
Our results highlight the strengths and weaknesses of each architecture in the different scenarios. 

\subsection{Testing} \label{testing}

Each experiment uses data pulled from the Alexa Top 1 Million list \cite{alextop1mil} and DGArchive \cite{dgarchive}. The only three available dictionary DGA families are considered: \texttt{gozi}, \texttt{matsnu}, and  \texttt{suppobox}. For model training and validation, the data is always separated into three sets: training, testing, and holdout. Training and testing are used at every epoch to see if early stopping should occur, preventing overfitting. The results for each metric, listed in Section \ref{Results}, are from applying the model to the holdout set.

\subsubsection{Testing Classification}  \label{testingclassification}

The first test evaluates how the model performs for binary classification between benign (negatives) and dictionary DGA domains (positives). With a balanced dataset, 80\% was used for training the model. The remaining 20\% (approximately 60,000 domains) was randomly sampled to use for testing and holdout: 50,000 domains for testing the model at each epoch and 10,000 domains for validating the model after training was completed. All training, testing, and validation data sets contained an approximately equal number of positives and negatives. 

\subsubsection{Testing Generalisability}

This test evaluates how the model generalises to unseen dictionary DGAs. For this, three trials are created from the data sorted by dictionary DGA family. Each trial takes two of the families for training and splits the third over testing and holdout. For example, one variant uses \texttt{matsnu} and \texttt{suppobox} domains to train the model while evaluating the model's performance using \texttt{gozi} domains. This paper is the first to test DGA detection models in this way. 
% In some cases different DGA families generate the same domains; these are not removed from the testing set to mimic potential events in the real world.

\subsubsection{Testing Time-based Resiliency}

DGAs have been found to evolve over time, varying their generation algorithms slightly or using entirely new dictionaries \cite{kumar2019deploymentsystem}. While our tests for generalisability highlight some of the deep learning models' ability to classify alterations in the dictionary DGA, they are limited by our scope of sampling in 2016-17. 

To test detection system's resiliency on future versions of dictionary DGA domains, we evaluate our models trained on data from 2016-17 with DGA samples from November 2019. 
% These samples contain domains from current versions of \texttt{gozi}, \texttt{matsnu}, and  \texttt{suppobox}. Extra domains from other DGAs are also included in the results as they empirically follow dictionary DGA patterns and reveal model behaviour unseen in other tests. 
Models trained on all three dictionary DGA families are applied to this dataset. 

\subsection{Implementation of Deep Learning Models}

%%%%%%%%%%%%%
\begin{table*}[t]
 \caption{  
Samples of identified dictionary DGA domains with the top 50 scores from our holdout set from each component model (LSTM and CNN) and our hybrid, Bilbo. Blue are domains seen initially in the LSTM's top 50 samples and then in Bilbo's top 50 samples. Same for the yellow domains, but seen in the CNN samples and then in Bilbo's samples. Orange is a domain that appeared in different ranks within all three models.
\label{fig:model-examples}
}
% \textcolor{red}{\hl{foo}}

\begin{center}
\begin{tabular} [width=0.9\columnwidth] { c c c c}
	\toprule
	LSTM & CNN & $\rightarrow$ & Bilbo \\ 
	\midrule
    \texttt{alexandreaannabeth} &  \colorbox{orange}{\texttt{themshallsubjectbeenathe}} & & \colorbox{yellow}{\texttt{deforrestharrelson}} \\ 
\texttt{withinlaughter} & \colorbox{yellow}{\texttt{deforrestharrelson}} & & \colorbox{cyan}{\texttt{gwendolynchristopherson}} \\ 
\colorbox{cyan}{\texttt{gwendolynchristopherson}} & \texttt{harrietteharrelson} & & \colorbox{yellow}{\texttt{kimberleepatterson}}\\
\colorbox{cyan}{\texttt{kimberleighharoldson}} & \colorbox{yellow}{\texttt{kimberleepatterson}} & & \colorbox{cyan}{\texttt{kimberleighharoldson}}\\
\colorbox{orange}{\texttt{themshallsubjectbeenathe}} & \texttt{severallaughter} & & \texttt{oughtinterrupttrotheth}\\
\texttt{walkjuly} & \texttt{decemberheight} & & \colorbox{orange}{\texttt{themshallsubjectbeenathe}}\\
\bottomrule
\end{tabular}
\end{center}
\end{table*}
%%%%%%%%%%%%%

Deep learning models take numerical sequences as input. Thus, every domain string is encoded as an array of integers and then padded with zeros to ensure that all inputs are of the same size. Each Unicode character is mapped to an integer through a constructed list of 40 valid domain-name characters. For example, ``google'' would be converted to \texttt{[7, 15, 15, 7, 12, 5]} and padded with zeros at the beginning to get all inputs up to our maximum length of a domain string: 63 characters. Our final input is \texttt{[0, 0, ..., 0, 7, 15, 15, 7, 12, 5]}. During initial iterations, we confirmed that padding the end of the sequence made no difference when compared with padding the beginning of the sequence. Rather than a common embedding for all deep learning models, the embedding is learned by the model during training. The outputs from each deep learning model is a score, a single float between zero and one. This value indicates the model's confidence that the domain was generated by a dictionary DGA.

We compare our main model, Bilbo, against four models adapted from state-of-the-art architectures: a single layer ANN, a CNN, an LSTM, and MIT's Hybrid \cite{vosoughi2016tweet2vec, yu2018character}. The code for each model is in Appendix A. As mentioned in Section \ref{relatedwork}, deep learning models have frequently been shown to outperform feature-based approaches for DGA detection and are capable of millisecond scoring speeds. Because of these ideal characteristics for a dictionary DGA detection system, Bilbo is only compared to other deep learning architectures.
 
 All models were built in Keras \cite{keras} using the TensorFlow \cite{tensorflow} backend on a MacBook Pro to convey the ease for model retraining and that models can be deployed on smaller cloud servers. 
Each model is trained three times for ten epochs with a batch size of 512. 

%%%%%%%%%%%%%%%%%%%%%%%%%%%%%
\begin{figure}[t]
\begin{center}
\includegraphics[width=0.9\columnwidth]{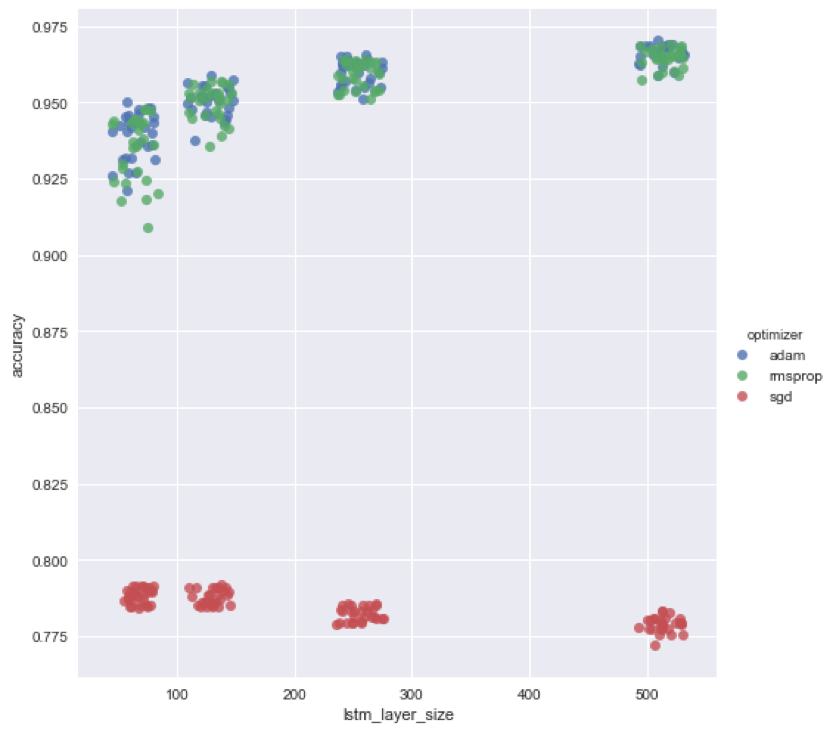}
\caption{
Graph of hyperparameter grid search used to inform decisions on the LSTM architecture. The LSTM layer size and optimiser are compared for accuracy on the test set, demonstrating improved performance using larger networks and either the \texttt{adam} or \texttt{rmsprop} optimiser.
\label{fig:lstm-krang}
\vspace{-1em}
}
\end{center}
\end{figure}
%%%%%%%%%%%%%%%%%%%%%%%%%%%%%

\subsubsection{Artificial Neural Network (ANN)}

This fundamental model architecture underlies both the CNN and LSTM. As a baseline for this study, similar to Yu et al. \cite{yu2018character}, we include a single-layer ANN with 100 neurons in its hidden layer during our testing and consideration. 
This architecture is also included within Bilbo as the conjoining layer for the parallel CNN and the LSTM component architectures. 

\subsubsection{Long Short-Term Memory (LSTM) Network}

This architecture is a slight adaptation on the LSTM used by Woodbridge et al. \cite{woodbridge2016lstm}. Because it was tuned for a slightly different task, we re-evaluated some of its hyperparameters.
From our automated grid search of hyperparameters, as shown in Figure \ref{fig:lstm-krang}, it was clear that increasing LSTM layer size improved our accuracy on the testing set for generic binary classification. We found that an LSTM layer of 256 nodes provided us with the highest accuracy on the testing dataset without loss to its performance in real-time deployments.
The only alterations to the original model were the input parameters to match our standard across models and doubling the size of the LSTM layer. This is the same architecture implemented as a component within Bilbo.

% FIXME Why RMSProp perform so bad?

\subsubsection{Convolutional Neural Network (CNN)}

% The CNN, while learning fewer parameters than the LSTM, requires more complex design. 
We followed Saxe et al.'s parallel convolution structure \cite{saxe2017expose} to compare with state-of-the-art with a CNN. 
After testing a variety of filter sizes individually, combinations of various filters were also analysed to
find the best architecture for our task. Based LCS analysis for each family, the majority of substrings within dictionary DGAs appeared to be within the range of two to six characters.
This model's final architecture includes five different sizes (2-6 characters) of convolutions, 60 filters of each length with a stride of one character, and pooling later concatenated to provide a vast amount of information towards the final score. This architecture balances the model complexity against the prediction accuracy on our training set. 

\subsubsection{Bilbo}

Our initial results with the individual LSTM and CNN, as seen in Table \ref{fig:model-examples}, indicated each model was learning relevant but distinct characteristics for accurate identification of dictionary DGAs. 
% We hypothesized that a hybrid model that combines the outputs of both the LSTM and CNN would be able to incorporate the specialised feature generation inherent in both model architectures and improve accuracy on binary classification tasks.  Table \ref{fig:model-examples} empirically motivates our hybrid approach, demonstrating that Bilbo correctly classifies domains that were flagged by both the LSTM and CNN when trained as standalone models.
% Notice the consistency of top domains our LSTM and CNN carried through to the hybrid.
Bilbo's architecture "bags" the extracted features from the LSTM and CNN with a hidden layer of 100 nodes, from which a final prediction is rendered. This hybrid model learns to balance the features extracted by both the LSTM and CNN. The same architectures described previously for the individual ANN, LSTM, and CNN are combined to form Bilbo. This model is the first parallel usage of a CNN and LSTM hybrid for DGA detection.

\subsubsection{MIT Hybrid Model}

Based on the original encoder-decoder model presented by MIT \cite{vosoughi2016tweet2vec},
several recent publications have adapted this CNN-LSTM hybrid model to DGA classification \cite{yu2018character, sivaguru2018evaluation, mohan2018spoof}.
Unlike our model, this uses the CNN convolutions to feed inputs into an LSTM. The MIT hybrid architecture adapted by Yu et al. \cite{yu2018character} is another benchmark during testing. Comparing Bilbo's parallel usage of a CNN and an LSTM to this model demonstrates the significance of our parallel architecture in binary classification of dictionary DGAs.

Their single convolutional layer consists of 128 one-dimensional filters, each three characters long with a stride of one. This is fed into a Max Pooling layer before a 64-node LSTM. This model contains no drop out and relies on a single sigmoid to flatten the results to a single score. 

\subsection{Metrics for Comparison}

Considering real-world applications for DGA detection, a balance between incorrect domains and lack of confidence for true dictionary DGA domains must be found.
To help measure each model's performance for this, three core metrics are calculated to summarise common metrics used in machine learning research. The first is the area under the receiver operating characteristic (ROC) curve (AUC), which measures the model's ability to detect true positives as a function of the false positive rate. Maximising AUC means improving labelling of both positive and negative samples. The second is accuracy; how well the model scored positive and negative labels our of all samples in the holdout set. Finally, the $F_1$ score is the harmonic mean of precision and recall, giving insight to the context of true positive labels within the holdout set.

Using abbreviations for true positive (TP), true negative (TN), false positive (FP), false negative (FN), true positive rate (TPR), and false positive rate (FPR), these are computed in the following ways:  

\begin{displaymath}
  Precision = \dfrac{\sum TP}{\sum TP + \sum FP}
\end{displaymath}

\begin{displaymath}
  Recall = \dfrac{\sum TP}{\sum TP + \sum FN}
\end{displaymath}

\begin{displaymath}
  F_1 = 2 * \dfrac{Precision * Recall}{Precision + Recall}
\end{displaymath}

\begin{displaymath}
  Accuracy  %= \dfrac{\sum TP + \sum TN}{Total Samples} 
  = \dfrac{\sum TP + \sum TN}{\sum TP + \sum FP + \sum TN + \sum FN}
\end{displaymath}

\begin{displaymath}
  TPR = \dfrac{\sum TP}{\sum TP + \sum FN} 
\end{displaymath}

\begin{displaymath}
  FPR = \dfrac{\sum FP}{\sum FP + \sum TN}
\end{displaymath}

Consistency of the core metrics in every setting is key to finding the best performer while evaluating models on labelled data. To quantify this, the core metrics are treated as assessment questions: one point of consistency is awarded to each of the top three models within every core metric. The model with the most points across testing classification and generalisability is deemed the most consistent performing model.

\section{Results} \label{Results}

In this section, we elaborate on the values of metrics from each model resulting from each test. The threshold for a label for every test was $0.5$.
Overall, the priority is to accurately apply both positive and negative labels to the dataset. From these tests, the model with the best consistency score is viewed as the best for deploying into real world settings.

%%%%%%%%%%%%%
\begin{table*} [t]
\caption{
 (Testing Classification) Comparing the results of five different deep learning architectures for binary dictionary-DGA classification. The labelled training and testing set are composed of a random selection from all three dictionary DGA families. The best of each column is in bold.
\label{table:test-classification}
\vspace{-1em}
}
\begin{center}
\begin{tabular}[width=0.9\textwidth] { c  c  c  c  c c c c} 
 \toprule
 Model & Recall & Precision & $F_1$ Score & TPR & FPR & AUC & Accuracy \\
 \midrule
 ANN & 0.9077 & 0.8250 & 0.8644 & 0.8250 & 0.1953 & 0.9290 & 0.8566 \\ 
 CNN & 0.9730 & 0.9473 & 0.9600 & 0.9473 & 0.0545 & 0.9919 & 0.9593 \\
 LSTM & 0.9675 & 0.9627 & 0.9651 & 0.9627 & 0.0370 & 0.9932 & 0.9653 \\
 MIT & 0.9583 & \textbf{0.9710} & 0.9646 & \textbf{0.9710} & \textbf{0.0282} & \textbf{0.9946} & 0.9651 \\
%  \midrule
 Bilbo & \textbf{0.9766} & 0.9557 & \textbf{0.9660} & 0.9557 & 0.0454 & 0.9944 & \textbf{0.9656} \\
 \bottomrule
\end{tabular}

\end{center}
\end{table*}
%%%%%%%%%%%%%

%%%%%%%%%%%%%
\begin{table*} [t]
\begin{center}
\caption{
 (Testing Generalisability) Comparing the results of deep learning architecture for generalisability of dictionary DGA classification. As documented for each trial, the top row lists the training dictionary DGA families with an arrow going to the testing family. The best of each column is in bold.
\label{table:test-generalisability-results}
\vspace{-1em}
}
\begin{tabular}[width=0.9\textwidth] { c  c   c c c  c  c c c  c  c c c} 
 \toprule
  Model &
  & \multicolumn{3}{c }{\texttt{matsnu} + \texttt{suppobox} $\rightarrow$ \texttt{gozi}} & 
  & \multicolumn{3}{c }{\texttt{matsnu} + \texttt{gozi} $\rightarrow$ \texttt{suppobox}} & 
  & \multicolumn{3}{c }{\texttt{gozi} + \texttt{suppobox} $\rightarrow$ \texttt{matsnu}} \\ 
 &
 & AUC & $F_1$ & Accuracy &
 & AUC & $F_1$ & Accuracy &
 & AUC & $F_1$ & Accuracy \\ 
 \midrule
 ANN && 0.8347 & \textbf{0.7465} & \textbf{0.7514} &
        & 0.7728 & 0.5858 & 0.6665 &
        & 0.7805 & \textbf{0.7033} & \textbf{0.7230} \\
 CNN && 0.9129 & 0.5881 & 0.6954 &
        & \textbf{0.8140} & 0.5909 & \textbf{0.6855} &
        & 0.8180 & 0.5038 & 0.6537 \\ 
LSTM && 0.8797 & 0.4066 & 0.6149 &
        & 0.7556 & \textbf{0.6010} & 0.6739 &
        & 0.8414 & 0.3189 & 0.5862 \\ 
MIT && 0.8971 & 0.3923 & 0.6103 & 
        & 0.7616 & 0.5379 & 0.6612 &
        & \textbf{0.8439} & 0.4600 & 0.6336 \\ 
% \midrule
%  Bilbo && \textbf{0.9137} & \textbf{0.5357} & \textbf{0.6708} &
%         & \textbf{0.8032} & 0.5660 & \textbf{0.6729} &
%         & \textbf{0.8309} & 0.4218 & 0.6217 \\ 
 Bilbo && \textbf{0.9137} & 0.5357 & 0.6708 &
        & 0.8032 & 0.5660 & 0.6729 &
        & 0.8309 & 0.4218 & 0.6217 \\ 
 \bottomrule
\end{tabular}

\end{center}
\end{table*}

%%%%%%%%%%%%%

%%%%%%%%%%%%%
\begin{table} [t]
\caption{
 (Testing Time-based Resiliency) Ratio correct at scoring dictionary DGAs from each family within the sample from November 2019. The best of each column is in bold.
\label{table:test-time-resiliency}
\vspace{-1em}
}
\begin{tabular}[width=0.9\columnwidth] { c  c  c  c}
\toprule
 Fully-Trained Model & \texttt{matsnu} & \texttt{suppobox} & \texttt{gozi} \\ %& Extra \\
 \midrule
 ANN & 0.8746 & 0.9204 & 0.9517 \\ %& \textbf{0.6307} \\ 
 CNN & 0.8732 & 0.9962 & \textbf{0.9585} \\ %& 0.3286\\ 
 LSTM & 0.8936 & 0.9522 & 0.9426 \\ % & 0.4906 \\ 
MIT & 0.8790 & \textbf{0.9976} & 0.9386 \\ %& 0.3213\\ 
%  \midrule
 Bilbo & \textbf{0.9023} & 0.9962 & 0.9472 \\ %& 0.4116 \\ 
 \bottomrule
\end{tabular}
\end{table}
%%%%%%%%%%%%%%%%%%%%%%%%%%%%%%%%%%%%%%%%%%%%%%%%%%%%

%%%%%%%%%%%%%
\begin{table} [t]
\caption{
 Consistency scores from each of the tests (1 = classification, 2 = generalisability, 3 = time-based resiliency) and the overall result. Calculated by counting the number of times each model was top three for a core metric. The best of each column is in bold.
\label{table:consistency-scores}
\vspace{-1em}
}
\begin{tabular}[width=0.9\columnwidth] { c  c  c  c c c c c}
\toprule
 Model & Test 1 & $+$ & Test 2 & $+$ & Test 3  & $=$ & Overall \\ 
    %   & Consistency && Consistency && Consistency \\
 \midrule
 ANN & 0 && 6 && 0 && 6 \\
 CNN & 0 && \textbf{8} && \textbf{1} && 9 \\
 LSTM & \textbf{3} && 3 && 0 && 6 \\
MIT & \textbf{3} && 4 && \textbf{1} && 8 \\ 
%  \midrule
 Bilbo & \textbf{3} && 6 && \textbf{1} && \textbf{10} \\
 \bottomrule
\end{tabular}
\end{table}
%%%%%%%%%%%%%%%%%%%%%%%%%%%%%%%%%%%%%%%%%%%%%%%%%%%%

\subsection{Results of Testing Classification}

% All five models have been trained on 80\% of all dictionary DGA samples and an equal amount of benign samples. Then each is tested and validated with the other 20\% of the data, again equal positive and negative labels. Threshold of $0.5$ was used for each of the metric calculations.
The values for the metrics from this test are provided in Table \ref{table:test-classification}. 
In this test, the ANN is significantly worse than the specialised deep learning models in every metric, according to a student t-test with 95\% confidence on the all collected results.
The ANN's FPR of $0.1953$ is almost a whole magnitude worse than MIT's FPR, which was the best. 

The CNN and LSTM are statistically similar in all metrics with the LSTM outperforming the CNN in most precision, TPR, and FPR. This is due to the imbalance between the dictionary DGA families, with \texttt{suppobox} comprising of about 78\% of the malicious samples. During our substring analysis, we found that \texttt{suppobox} contained the longest substrings, revealing that models which learn the long sequence of \texttt{suppobox}'s dictionary words would have an advantage when classifying the majority of dictionary DGA domains. The LSTM is designed to learn sequential relationships between characters rather than subsets of characters like the CNN. This is why the LSTM beats the CNN and, as shown in Table \ref{table:consistency-scores}, is a consistent leader in the core metrics.

Both MIT's hybrid model and Bilbo perform the best across all metrics. The difference between the two is insignificant in all metrics, differing less than $0.01$ for the $F_1$ score, Accuracy, and AUC.
This near identical performance is similar to the LSTM and CNN comparison earlier. There is also a pattern in most of the metrics that when the CNN is better than the LSTM, Bilbo is better than the MIT model and vice versa. MIT's parameters are mostly dedicated to the LSTM layer, explaining the similar performance between the two models.

Bilbo consistently performs between or better than its component models in all metrics by regularising the performance of the LSTM and CNN with an ANN, displaying the expected results of our parallel architecture. In the empirical analysis of the results, the top scoring domains from both the CNN and LSTM were present in the final scoring of Bilbo as expected. 

The difference between the deep learning models, excluding the ANN, in this test is very small. Given a domain name, they are all successful at labelling dictionary DGA domains from benign domains after learning from three diverse dictionary DGA families. The consistency scores for this test place the LSTM model, the MIT model, and Bilbo as the best performers.

\subsection{Results of Testing Generalisability}

As presented in Table \ref{table:test-generalisability-results}, the metrics have been limited to three core metrics to maximise for best overall performance. A model's AUC indicates the model's likelihood of correctly classifying a sample as a positive or negative. The $F_1$ score conveys how well the model correctly labels dictionary DGA domains with regard to those that should be or were labelled. Accuracy states how well the model labelled the data within this particular holdout set.

The values for the core metrics were not expected to surpass $0.9$ due to the differences between each dictionary DGA family. Analysis of each family's LCS and the JW scores between families not depicted in this paper stated some families overlap more with one family than another. This dependence influences each model's performance by limiting its ability to generalise unless certain families have been seen before.
Hence the values across this table are lower than in Table \ref{table:test-classification}.

The ANN outperforms the other models in this task with higher core metrics in two of the trials. However, it also only surpasses the other models when \texttt{matsnu} or \texttt{gozi} are part of the training set. Figure \ref{fig:chord-counts} depicts a large overlap in their LCS. This could explain what the ANN is able to learn for better performance on new DGAs when either \texttt{matsnu} or \texttt{gozi} is in the training set and the other is in the testing set. 

The next most consistent performer in this test is the CNN. Its training on smaller character windows allows it to excel when applied to new dictionary DGAs. Based on earlier data analysis, the most frequent LCS in every family were three to four characters and typically overlapped. The large overlap in LCS between \texttt{matsnu} and \texttt{gozi} reinforce these short substrings, explaining why the CNN outperforms others when both \texttt{matsnu} and \texttt{gozi} are in the training set.

% There is an approximately $0.02$ difference between the CNN and Bilbo for most of the core metrics. However, unlike Bilbo, the CNN was never a top performer in the previous classification test.
%%%%%%%%%%%%%%%%%%%%%%%%%%
% Looking at the AUC for each model in each trial, Bilbo is consistently in the top three: it has the best AUC when testing on \texttt{gozi}, the second best when testing on \texttt{suppobox}, and the third best when testing on \texttt{matsnu}. Returning to Figure \ref{fig:chord-counts} from our data analysis, it shows that a large portion of substrings from Matsnu and Gozi overlap. What is not fully conveyed through that figure is the majority of those substrings are 3-5 characters long. Several of these substrings could be within the same domain and could combine to form the longer strings found in \texttt{suppobox}, explaining why the CNN does best on the trial testing on \texttt{suppobox}.

% Based on the MIT's hybrid architecture using an LSTM for learning sequences from convolutional filters, it is apparent that the LSTM architecture heavily influences the MIT model's performance. The scores between the LSTM and MIT models are very similar, with the convolutions providing a slight advantage. With \texttt{suppobox} in the training set, the number of training samples increases by over 100,000; without it, the training set contains roughly 40,000 samples. Learning the sequences and increasing the training set size improves the LSTM and MIT's AUC to 0.01 above Bilbo's AUC. 

\subsection{Results of Testing Time-based Resiliency}

% The final test is on a single day's worth of recent domain samples from each of the dictionary DGA families already considered and a few empirically selected domains following a similar expected dictionary DGA pattern but from other malware families. These additional samples are labeled as ``Extra'' in Table \ref{table:test-time-resiliency} with the other results and are from DGA families not typically dictionary-based. Listed are the ratios of true positives (above or equal to 0.5 threshold) out of the total number of samples for that malware family. Total samples from each malware family are as follows: 1325 from \texttt{gozi}, 686 from \texttt{matsnu}, 4257 from \texttt{suppobox}, and 639 from \texttt{banjori} and \texttt{nymaim}, labelled as ``Extra''.
The final test is on a single day's worth of recent domain samples from each of the dictionary DGA families already considered. Listed are the ratios of true positives out of the total number of samples for that dictionary DGA family. Total samples for each family are as follows: 1325 from \texttt{gozi}, 686 from \texttt{matsnu}, and 4257 from \texttt{suppobox}.

Using all of the trained models from the classification test, the average scores are listed. The results are close between all model architectures and, when averaged, are close to the accuracy seen during testing. As for the relative decrease in accuracy for \texttt{matsnu} and \texttt{gozi}, this is due to the class imbalance between the dictionary DGA families in the dataset. 
% Unless the dataset is altered to cover more balanced number of samples from each family, it will need to be updated frequently with new labeled data. 
Regardless of which model selected for deployment, it will need to be updated frequently with new labelled data whenever trusted and available to increase this accuracy on future dictionary DGA domains.

Throughout all of these tests, each state-of-the-art deep learning model achieves top metrics. To determine which is the best, we consider the application environment the model is to be deployed in and its need for a consistent well-performing model. After aggregating the consistency points for the top performers from every core metric in each test and trial, presented in Table \ref{table:consistency-scores}, Bilbo is found to be the most consistent and capable model for deploying within real-world dictionary DGA detection systems.

\section{Real-World Deployment} \label{real-time}

Once Bilbo was trained, tested, and validated using open source data from the Alexa Top 1 Million \cite{alextop1mil} and DGArchive \cite{dgarchive}, we evaluated performance in a live system. We deployed the model on a cluster of servers to be queried by a data pipeline and applied the model to live network traffic from a large enterprise.  

\subsection{Implementation at the Corporate Level}

%%%%%%%%%%%%%%%%%%%%%%%%%%%%%%%%%%%%%%%%
\begin{figure*}[t]
\begin{center}
\includegraphics[width=0.9\textwidth]{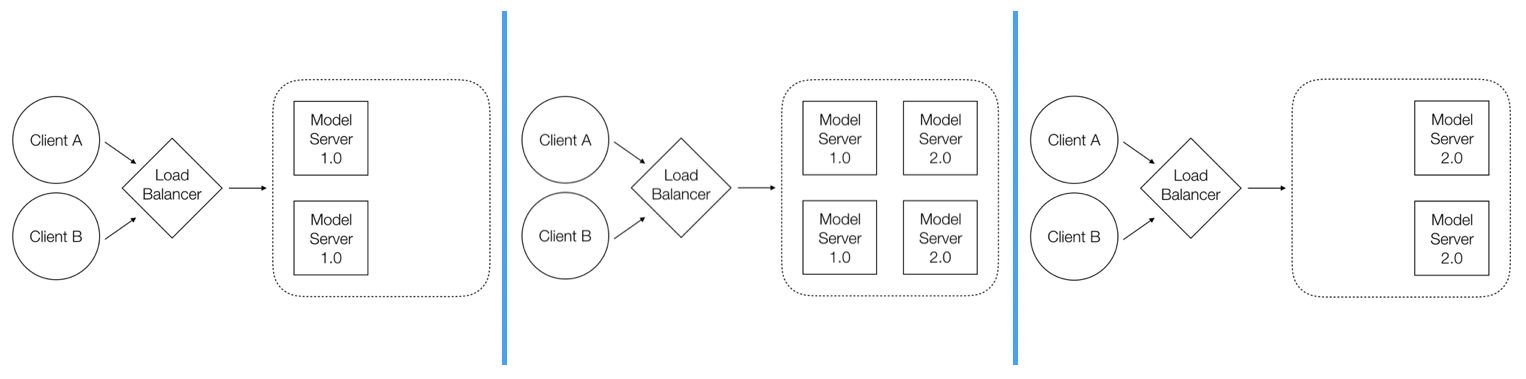}
\caption{
Three stages of the updating process for our Model as a Service (MAAS) Architecture for model deployments to be accessible to SIEM and other client systems. From left to right, the first stage shows clients interacting through the load balancer with old model servers. To update the servers, we spin up new model servers with the latest version and confirm production readiness before attaching them to the load balancer. Finally, the old model servers are deleted, leaving the new model servers in their place. At no point during this process will the clients be unable to receive scores from our models.
\label{fig:maas-arch}
\vspace{-1em}
}
\end{center}
\end{figure*}
%%%%%%%%%%%%%%%%%%%%%%%%%%%%%%%%%%%%%%%%

Within corporate environments, a large security information and event management (SIEM) system is typically used to centralise and process relevant data sources. Security analysts use the SIEM for their daily work to investigate suspicious activity within their environment. The data they view is limited by a series of filters and joins they apply on various datasets. 

To productionise Bilbo in a high-throughput environment generating hundreds of domains per second, we developed a model as a service framework. This framework promotes scalability, modularity, and ease of maintenance. Client systems processing domain names, such as the SIEM, make requests of the model servers to receive scores on new domains. This communication is performed using gRPC, Google's library for remote procedure calls \cite{grpc},
which was selected for its speed over methods like REST (Representational State Transfer). The communication from client to server is language-agnostic, allowing a client written in Java or Scala to interface seamlessly with our Python model.

A load balancer manages traffic to the model servers and only the load balancer endpoint is exposed to the client. This allows multiple clients to reach out to a single location in order to receive scores from the model. Any number of model servers can run behind the load balancer, but these details are abstracted away from the clients, who only interface with the load balancer endpoint. This allows us to increase and decrease the size of the model server cluster in response to changing without interrupting service; such scaling can be configured to take place automatically in response to metrics like CPU utilisation.

While our model does not learn inline, its predictions, combined with a ground truth label provided by an analyst, can be used to retrain the model, allowing it to learn from mistakes and improve its predictive power. Thus, we need to be able to deploy a retrained model frequently and with low overhead. Since the model server cluster is behind a load balancer, we can make this change without shutting down the service. We simply put additional model servers (running the newest model) behind the load balancer, and, once they have been confirmed to run successfully, remove the model servers running an outdated version. The model update process can be seen in Figure \ref{fig:maas-arch}. Along with their scores, the model servers return the version of the model that they are running; this is helpful in evaluating our models over time and in distinguishing between models during the brief overlap period when two versions of the model are running behind the load balancer.

Several key design decisions allow us to handle requests to the service at very large scale. While gRPC minimises network latency by allowing bi-directional streaming between the client and server, the calls to our service are still time-intensive, so we built in a bloom filter caching mechanism on the client side to avoid this bottleneck. This more intelligent client only reaches out to the server if it receives a domain that it has not recently seen before. Our analysis of domain traffic revealed that only 15\% of domains are unique in an hour of traffic; this optimisation dramatically reduces the workload of our model server cluster.

We evaluated Bilbo based on its processing capacity and its findings, as seen below.
Our initial prototype consisted of a single client reaching out to a load balancer with a single server in the cloud. With an unoptimized compilation of Tensorflow for our backend, the fastest scoring averaged to approximately 10 ms per record, increasing linearly with an increasing number of requests. If we anticipate 1000 domains per second, our model only needs to be hosted on 10 servers. On a Cloud service such as Amazon Web Services, we can keep a ten-node cluster running for less than fifty cents (USD) per hour.

\subsection{Results in Enterprise Traffic}

% For further model benchmarking, we used four hours of network traffic from a large enterprise. This dataset is not publicly available, but relevant findings are discussed in section \ref{Validation}. The network traffic logs were parsed to extract domains and each domain was scored by our model. We also used VirusTotal (accessed November 2017) and other third-party security tools to analyze extracted domains and provide a comparison to the model's results.

For further model performance testing, Bilbo is evaluated on real-world network traffic. Randomly selecting one window of traffic from August 14\textsuperscript{th}, 2017, and another window of traffic from November 15\textsuperscript{th}, 2017\footnote{These were recent dates when the model was initially developed for dictionary DGA detection.}. Each network sample set contains domain names over a two-hour period. After parsing the domain names from the URLs in the logs, the August and November data contained 20,000 and 45,000 unique domains, respectively.

Since we lack ground truth for the domains in our captured samples to validate our results, we pulled in additional information for each domain. First, we included the action decision of the proxy, which denies domains that are known to be malicious. Second, we added scores from VirusTotal \cite{virustotal}, a site that aggregates blacklists to provide reputation scores for domains and is commonly used by security analysts for evaluation of domains (accessed November, 2017). Note that both the proxy and VirusTotal are imperfect since they are unaware of malicious content related to a domain until thorough analysis has been performed, which can take many weeks \cite{leverlustrum}. We cross-referenced the high scores from our model with the results from the proxy and VirusTotal to perform a basic investigation.

Feeding our model only the domain names, we discovered a series of domains with similar naming patterns: 
\begin{itemize}
\item cot.attacksspaghetti[.]com/affs
\item kqw.rediscussedcudgels[.]com/affs
\item psl.substratumfilter[.]com/affs
\item dot.masticationlamest[.]com/affs
\end{itemize}
At a glance, these domains follow an algorithmic pattern of three characters, two words, and the ``/affs'' ending, making them strong candidate dictionary-based DGA domains. Upon further examination, all of these domains were queried by the same machine, which, prior to our discovery, had been deactivated due to complaints of incredibly sluggish performance. This is highly suggestive of malware activity using a dictionary DGA network.

\begin{figure}[t]
\begin{center}
\includegraphics[width=0.9\columnwidth]{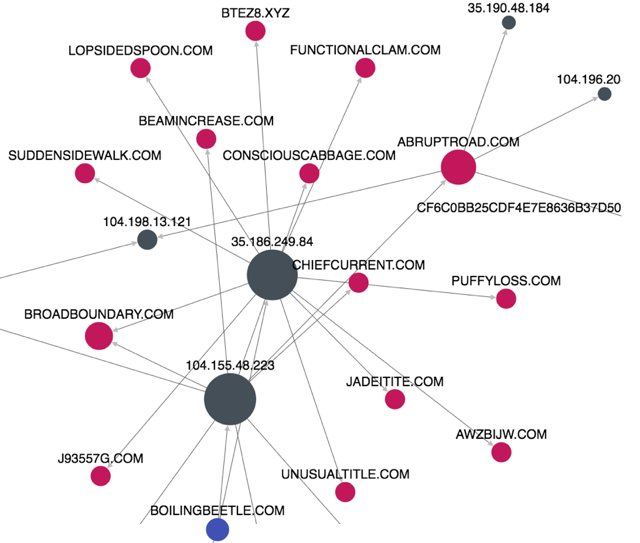}
\caption{
(Created December 2017) ThreatCrowd network graph of the domain ``boilingbeetle'' discovered in enterprise proxy traffic by the ensemble model. This domain is connected through select IP addresses to other domains of similar structure, in the pattern of a command and control network. 
\label{fig:threatcrowd}
\vspace{-1em}
}
\end{center}
\end{figure}

Additionally, we found four domains, each representing distinct suspicious networks matching the expected pattern for dictionary-based DGA C\&C hubs:
\begin{itemize}
\item boilingbeetle
\item silkenthreadiness
\item mountaintail
\item nervoussummer
\end{itemize}
Each of these networks, when visualised by ThreatCrowd, a crowd-sourced network analysis ``system for finding and researching artefacts relating to cyber threats'' \cite{threatcrowdref}, are shown to be comprised of domains that are made up of two or more unrelated words, all resolving to the same IP address, in the pattern of a domain-fluxing dictionary DGA. The ``boilingbeetle'' network is shown in Figure \ref{fig:threatcrowd}. These domains and their related networks were not flagged by the online blacklists used by VirusTotal; only some of the domains within each network were blocked by the proxy.

Further investigation noted that these networks are for advertisement traffic, indicating that dictionary DGA techniques are being used to bypass ad-blocker mechanisms. Although not apparently malicious, these five discoveries of dictionary-based DGA from potential malware, found in only a few hours of proxy log data, demonstrates that our solution is able to flag relevant results in live traffic.

%%%%%%%%%%%%%%%%%%%%%%%%%%%%%%%%%%%%%%%%
%%%%%%%%%%%%%%%%%%%%%%%%%%%%%%%%%%%%%%%%

\section{Conclusions and Future Work}

In this paper, we present a parallel hybrid architecture named Bilbo, composed of an LSTM, a CNN, and an ANN, for dictionary DGA detection. Dictionary DGAs bypass most general, manually-defined DGA defences and are harder to detect due to their natural language characteristics. 
% We trained our model with an even split of empirically selected dictionary DGA families from DGArchive and random domains from the Alexa Top 1 Million list. 
Bilbo is compared to state-of-the-art deep learning models adapted for dictionary DGA classification and evaluated on consistency over AUC, Accuracy, and $F_1$ score. Overall, Bilbo is the most consistent and capable model available.

Bilbo was then applied to a large financial corporation's SIEM, providing inline predictions within a scalable framework to handle high-throughput network traffic. During investigations, our model's scores were used to filter data and flag suspicious activity for further analysis. 

When applied to several hours of live network logs, Bilbo successfully classified traffic matching the expected network pattern: a single IP address hosting several domain names that make no semantic sense and follow a trend of English words put together. Although the identified domains from the network logs were not botnets or worms reaching out to a C\&C, which are very rare, Bilbo was able to identify dictionary DGAs used by advertisement networks and other applications with potential malicious intent.

% At peak hours in a large enterprise, a SIEM processes thousands of domains per second, requiring a domain analysis system to make decisions in milliseconds. This time constraint prohibits techniques based on historical lookups, which have been successful in general DGA detection.
% Other alternatives for future exploration access other contextual features from a DNS query, however pursuit of these features may increase latency of our real-time analysis. 

Later improvements include the continued reduction of false positives and applying natural language processing (NLP) techniques. One method to reduce false positives would be to consider 
layering a generative model to determine if the input domain is similar to any data Bilbo has seen before. This could increase or decrease the score, or add another filter to alter a user's confidence in the score. 
Applicable NLP techniques detect anomalous word combinations in domains by scoring the likelihood words would be collocated. This could prove fruitful for DGA detection but heavily depends on the corpus for parsing out words and gathering initial collocation information to understand for a baseline of what is normal. 

\begin{acks}
Thank you to Capital One for the incredible opportunity to deploy a machine learning model developed for research into a live environment for evaluation. To Jason Trost, your mentorship and intellectual curiosity inspires everyone around you. We appreciate your and Capital One's support to publish our work as an academic paper after our talks in industry.

To the reviewers at our last attempted venues, thank you for the incredible feedback that greatly improved our analysis.
\end{acks}

\bibliographystyle{ACM-Reference-Format}
% \bibliography{bib/data_tools, bib/deeplearning, bib/domains}
\bibliography{bilbo-sigconf}

\appendix

\section{Model Architectures}

All code can be found at \href{https://github.com/jinxmirror13/bilbo-bagging-hybrid}{this Github repository:}. 

\href{https://github.com/jinxmirror13/bilbo-bagging-hybrid}{https://github.com/jinxmirror13/bilbo-bagging-hybrid}

\end{document}